\documentclass[prd,floatfix,nofootinbib,preprintnumbers]{revtex4}
\usepackage{epsfig,subfigure}
\usepackage{graphicx}
\usepackage{amssymb}

\begin{document}

%\draft
\newcommand{\be}{\begin{equation}}
\newcommand{\ee}{\end{equation}}
\newcommand{\bea}{\begin{eqnarray}}
\newcommand{\eea}{\end{eqnarray}}
\newcommand{\nnb}{\nonumber}
\renewcommand{\thefootnote}{\fnsymbol{footnote}}
%%%%%%%%%%%%%%%%%%%%%%%%%%%%%%%%%%%%%%%%%%%%%%%%%%%%%%%%%%%%%%%%%%%%%%%
\def\lsim{\raise0.3ex\hbox{$\;<$\kern-0.75em\raise-1.1ex\hbox{$\sim\;$}}}
\def\gsim{\raise0.3ex\hbox{$\;>$\kern-0.75em\raise-1.1ex\hbox{$\sim\;$}}}
\def\Frac#1#2{\frac{\displaystyle{#1}}{\displaystyle{#2}}}
\def\no{\nonumber\\}
\def\slash#1{\ooalign{\hfil/\hfil\crcr$#1$}}
\def\ep{\eta^{\prime}}
\def\susy{\mbox{\tiny SUSY}}
\def\sm{\mbox{\tiny SM}}
\def\pslash{\rlap{\hspace{0.02cm}/}{p}}
\def\qslash{\rlap{/}{q}}
\def\kslash{\rlap{\hspace{0.02cm}/}{k}}
\def\lslash{\rlap{\hspace{0.011cm}/}{\ell}}
\def\nslash{\rlap{\hspace{0.02cm}/}{n}}
\def\Pslash{\rlap{\hspace{0.065cm}/}{P}}
%%%%%%%%%%%%%
%\setlength{\topmargin}{-.75in} %this gives you a 1" topmargin.
%\setlength{\evensidemargin}{-0.5in}
%\setlength{\oddsidemargin}{-0.5in}
%\setlength{\textwidth}{7in} \setlength{\textheight}{9in}
\textheight      250mm  % height of text

%\twocolumn[\hsize\textwidth\columnwidth\hsize\csname@twocolumnfalse\endcsname
%\rightline{\small IPPP/03/11 \, \, DCPT/03/22}
\vskip0.5pc

\title{Implications for new physics from ${\bar B}^0\to \pi^0\pi^0$ and ${\bar B}^0\to {\bar K}^0K^0$}
\author{Jian-Feng Cheng$^a$, Yuan-Ning Gao$^a$,  Chao-Shang Huang$^b$, and Xiao-Hong Wu$^{c}$}
\affiliation{
 $^a$ Center for High Energy Physics, Tsinghua University,
             Beijing 100084,  China\\
 $^b$ Institute of Theoretical Physics, Academia Sinica, P. O. Box 2735,
             Beijing 100080,  China\\
 $^c$  School of Physics,  Korea Institute for Advanced Study, Seoul 130-722, Korea }

\begin{abstract}
We have analyzed the ${\bar B}^0\to \pi^0\pi^0$ puzzle in three
kinds of models beyond the standard model (SM). It is shown that
the minimal flavor violation (MFV) models, the minimal
supersymmetric standard model (MSSM), and the two Higgs doublet
models (2HDM) I and II can not give an explanation of the ${\bar
B}^0\to \pi^0\pi^0$ puzzle within $1 \sigma$ experimental bounds
and the model III 2HDM can explain the puzzle without a conflict
with other experimental measurements. If the constraint on
$C_{8g}$ from $b\to s\, g$ is not imposed, for all kinds of
insertions considered there are regions of parameter space, where
the scalar quark mass is larger (much larger) than the gluino mass
in the case of LR or RL (LL or RR), in which the puzzle can be
resolved within $1 \sigma$ experimental bounds.
\end{abstract}
\maketitle

The branching ratios of ${\bar B}^0$ decays into two pions have been
recently observed~\cite{hfag} : \begin{eqnarray}  Br({\bar B}^0\to
\pi^0\pi^0)&=&(1.45\pm 0.29)\times 10^{-6},
\\ \nnb Br({\bar B}^0\to \pi^+\pi^-)&=&(5.0\pm 0.4)\times 10^{-6}.\end{eqnarray}
The large branching ratio of the B decay into neutral pion final
states is unexpected. The decay amplitudes of ${\bar B}^0\to
\pi\pi$ can be generally parameterized as  \begin{eqnarray}
\sqrt{2}A({\bar B}^0\to \pi^0\pi^0)&=& T \left[ \left(
\frac{P}{T}-\frac{P_{EW}}{T}\right) e^{i\alpha}-\frac{C}{T}\right]
,\\ \nnb A({\bar B}^0\to \pi^+\pi^-)&=& -T \left[ 1+
\frac{P}{T}e^{i\alpha}\right],
\end{eqnarray}
where $T, C, P,$ and $P_{EW}$ are the tree, color-suppressed tree,
penguin, and electro-weak penguin (EWP) amplitudes respectively,
and $\alpha=\arg \left(-{\lambda_u\over\lambda_t}\right)$ is the
weak phase, where $\lambda_p=V_{pb}V^*_{pd} (p=u,c,t)$. In SM one
has the counting rules: the color-suppressed tree and penguin
amplitudes are suppressed by a factor of $\lambda$ ($\lambda\sim
0.22$ is the Wolfenstein parameter) and the EW penguin is
suppressed by a factor of $\lambda^2$, with respect to the tree
amplitude~\cite{tpc}. So one should expect by the naive counting
rules that the branching ratio of the B decay into neutral pion
final states is $O(\lambda^2)$ of that for charged pion final
states. However, the data (Eq.~(1)) indict that the former is
$O(\lambda)$ of the later. The observed branching ratio of ${\bar
B}^0\to \pi^0\pi^0=(1.45\pm 0.29)\times 10^{-6}$ is much larger
than the theoretical prediction, about $0.3\times 10^{-6}$, up to
the $\alpha_s$ order in the BBNS approach (QCDF)~\cite{bbns1,bn}
in SM. In Li et al's approach (PQCD) the leading order (LO)
prediction $\sim 10^{-7}$~\cite{luy} is the same order as that of
the QCDF prediction. In the recent paper~\cite{lms} the next
leading order (NLO) PQCD calculations have been carried out and
the results are that the $\pi K$ puzzle, the expected relation
$A_{CP}(B^{\pm}\to \pi^0 K^\pm)\approx A_{CP}(B^0\to \pi^\pm
K^\mp)$ disagreed significantly with the data, can be resolved but
the predicted branching ratio of ${\bar B}^0\to \pi^0\pi^0$ is
about $0.3\times 10^{-6}$ which is still much smaller than the
data, i.e., the $\pi^0\pi^0$ puzzle remains. If the large
branching ratio persists it could indicate new physics.% to resolve the puzzle.

Though ${\bar B}^0\to \pi^0\pi^0$ is not a pure-penguin process
and has the contributions from tree operators, the tree
contributions are of the order same as the penguin contributions
because of the almost completely cancellation between the two
terms in $C_2+C_1/N_c$ where $C_{1,2}$ are Wilson coefficients of
tree operators, so ${\bar B}^0\to \pi^0\pi^0$ is sensitive to new
physics. Therefore, it seems that a lot of new models beyond SM
could enhance the branching ratio and consequently resolve the
puzzle~\cite{yang}. However, any new model must simultaneously
give an explanation for the branching ratio of ${\bar B}^0\to
\bar{K^0}K^0$ since the two processes are closely related at quark
level: the flavor changing neutral current $b \rightarrow d$
transition controls ${\bar B}^0\to \bar{K^0}K^0$ and the same
transition gives significant contributions to ${\bar B}^0\to
\pi^0\pi^0$ which are of the order same as the tree contributions
in SM. Recently the branching ratio of ${\bar B}^0\to
\bar{K^0}K^0$ has been measured as $(0.96\pm 0.25)\times
10^{-6}$~\cite{hfag}, which is consistent with the prediction from
both the QCDF~\cite{bn} and PQCD approaches~\cite{cl}. Therefore,
new physics (NP) contributions must satisfy the condition that
they make Br of ${\bar B}^0\to \pi^0\pi^0$ enhanced but keep Br of
${\bar B}^0\to \bar{K^0}K^0$ basically unchanged, compared with
those in SM respectively, which will impose the significant limit
on NP models.

In the paper we search for new models beyond the SM which can
account for the data of branching ratios for both the ${\bar
B}^0\to \pi^0\pi^0$ and ${\bar B}^0\to \bar{K^0}K^0$ processes. To
be specific, we concentrate on the well-known three kinds of
models: the minimal flavor violation (MFV) models, the two Higgs
doublet models (2HDM) and the minimal supersymmetric standard
model (MSSM).

The effective Hamiltonian relevant for the two processes in the SM
can be expressed as\cite{bbns1}
\begin{eqnarray}\label{eff}
 {\cal H}_{\rm eff}^{SM} &=& \frac{G_F}{\sqrt2} \sum_{p=u,c} \!
   \lambda_p \bigg(C_1\,Q_1^p + C_2\,Q_2^p
   + \!\sum_{i=3,\dots, 10}\! C_i\,Q_i
   + C_{7\gamma}\,Q_{7\gamma}
   + C_{8g}\,Q_{8g}\bigg)+ \mbox{h.c.},
 \end{eqnarray}
where $\lambda_p=V_{pb} V^*_{pd}$, $Q_{1,2}$ and $Q_i (i=3,...10)$
are the tree and penguin operators respectively. Explicit forms
for $C_1, C_2, C_i,\,C_{7\gamma},\, C_{8g}$ and $Q_1^p, Q_2^p,
Q_i,\,Q_{7\gamma},\, Q_{8g}$ can be found, e.g., in
Ref.~\cite{bbns1}.

In the QCD factorization approach, the dominant contributions to
the decay amplitudes are given by:
\begin{eqnarray}\label{amp}
M({\bar B}^0\to \pi^0\pi^0) &=& {G_F\over\sqrt{2}} f_\pi F^{B\to \pi}m_B^2
\times {1\over \sqrt{2}} \sum_{p=u,c}\left[ a_2 \lambda_u
-(a_4^p+r^{\pi}a_6^p)\lambda_p
\right]\nonumber\\
M({\bar B}^0\to {\bar K}^0 K^0) &=& {G_F\over\sqrt{2}} f_K F^{B\to K}m_B^2 \times
\sum_{p=u,c}\left[ (a_4^p+r^K a_6^p)\lambda_p  \right], %\nonumber
\end{eqnarray}
where the definitions of the parameters $a_i$ and the chiral
enhancement factors $r^{\pi},\; r^K$ can be found in
Ref.~\cite{bbns1}. We take the values of running masses in the
$\overline{MS}$ scheme for light quarks such that $r^{\pi}=
r^K\equiv r$ hereafter. The electro-weak penguin and annihilation
contributions are neglected in above formula, which leads to the
$10\%$ theoretical uncertainty.

First we consider the MFV models. The MFV models beyond the SM
discussed in the paper mean a class of models in which the general
structure of flavor changing neutral current (FCNC) processes
present in the SM is preserved. In particular, all flavor
violating and CP-violating transitions are governed by the CKM
matrix and the only relevant local operators are the ones that are
relevant in the SM~\cite{mfv}. New parameters in the MFV models,
e.g., the masses of charginos, squarks, Higgs particles in the MFV
scenario of the MSSM, enter into Wilson coefficients of relevant
local operators. Therefore, in the MFV models the amplitudes of
the two decays are given same as Eq.~(\ref{amp}) (with values of
$a_i$ generally different from those in SM).

We can model-independently determine $|z|\equiv
|\sum_{p=u,c}\left[ (a_4^p+r a_6^p)\lambda_p \right]|$ from the
measured branching ratios of ${\bar B}^0\to \pi^0\pi^0$ and ${\bar
B}^0\to \bar{K^0}K^0$ and $a_2=( 0.0502 - 0.0689 i)\pm
(0.0025+0.0035 i)$ which comes from the known tree contributions.
The result is given in Fig. 1 where the $60\%$ theoretical
uncertainty (coming mainly from non perturbative parameters such
as form factors, distribution amplitudes and CKM matrix elements)
has been taken into account. There is a narrow region which can
simultaneously fit the data. However, assuming Wilson coefficients
of relevant operators, except the chromo-magnetic dipole operator,
change a little compared with SM, which is the case in the MFV
models, z in the region corresponds to \bea |C_{8g}(m_W)|\geqslant
2.6 \label{c8g}\eea which can not be reached in the MFV
models~\cite{algh,dgs}. That is, the MFV models are excluded
within $1 \sigma$ experimental bounds.
\begin{figure}\vspace{-2cm}
%\subfigure[ is for the model only with operators in SM.]{
%\begin{minipage}[t]{.48\textwidth}
%\epsfysize=4cm \vspace{-1.5cm} \centering
\includegraphics[width=8cm]{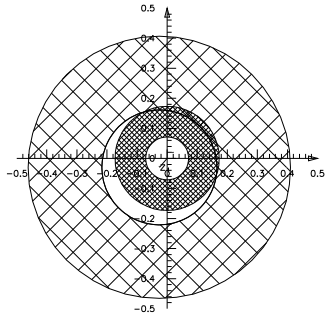} \vspace{-0.5cm}
\vspace{-1.5cm} \caption{The constraints on $z$ from the branching
ratios of ${\bar B}^0\to \pi^0\pi^0$ and ${\bar B}^0\to {\bar K}^0
K^0$. The big lattice denotes the constraint from ${\bar B}^0\to
\pi^0\pi^0$ and the small lattice for the constraint from ${\bar
B}^0\to {\bar K}^0 K^0$. }\label{independent}
\end{figure}

Next we consider models in which there are new operators in
addition to those in the SM, e.g., the 2HDM and MSSM. The
effective Hamiltonian in the 2HDM and MSSM can be written
as~\cite{ch,chw}
\begin{eqnarray}
\label{newh} {\cal H}_{\rm eff} &=& {\cal H}_{\rm eff}^{SM}+{\cal
H}_{\rm eff}^{new}, \\ \nnb
 {\cal H}_{\rm eff}^{new} &=& \frac{G_F}{\sqrt2} \sum_{p=u,c} \!
   V_{pb} V^*_{pd} \bigg(\!\sum_{i=11,\dots, 16}\![ C_i\,Q_i+ C_i^\prime\,Q_i^\prime]
   \nonumber \\&& + \!\sum_{i=3,\dots, 10}\!C_i^\prime\,Q_i^\prime
   + C_{7\gamma}^\prime\,Q_{7\gamma}^\prime
   + C_{8g}^\prime \,Q_{8g}^\prime \, \bigg) + \mbox{h.c.}, \,
\end{eqnarray}
where $Q_{i}^{(\prime)}$, i=11,...,16,  are the neutral Higgs
penguin operators and their explicit forms can be found in
Refs.~\cite{ch,chw} with the substitution $s\to d$. The primed
operators, the counterpart of the unprimed operators, are obtained
by replacing the chirality in the corresponding unprimed operators
with opposite ones.

>From the effective Hamiltonian, Eq. (\ref{newh}), it follows that
the main contributions to the decay amplitudes from the SM and new
physics are given by:
\begin{eqnarray}\label{newamp}
M({\bar B}^0\to \pi^0\pi^0) &=& {G_F\over\sqrt{2}} f_\pi F^{B\to \pi}m_B^2
\times {1\over \sqrt{2}} \sum_{p=u,c}\left[ a_2 \lambda_u -(a_4^p+ra_6^p)\lambda_p
\right]\nonumber\\
M({\bar B}^0\to {\bar K}^0 K^0) &=& {G_F\over\sqrt{2}} f_K F^{B\to K}m_B^2 \times
\sum_{p=u,c}\{ (a_4^p+ra_6^p)\lambda_p \nnb \\ && + {m_s\over m_b}
[h_1(C_{11}-C_{11}^\prime)+h_2
(C_{13}(\mu)-C_{13}^\prime(\mu))]\} %\right],
\end{eqnarray}
where we have set $m_d=0$. Due to the renormalization group
equation (RGE) running, Wilson coefficients $C_{i}$, i=14,15,16,
are related to $C_{13}$ and the known constants $h_{1,2}$
represent the running effects. The largest contributions to the
hadronic elements of the neutral Higgs penguin operators at the
$\alpha_s$ order arise from penguin contractions with $b$ quark in
the loop, which are the same for ${\bar B}^0\to \pi^0\pi^0$ and
${\bar B}^0\to \bar{K}^0 K^0$ and have been included in $a_4$
(see, for example, Ref.~\cite{cheng}). Therefore, although they
can enhance the branching ratio of ${\bar B}^0\to \pi^0\pi^0$,
they alone can not resolve the puzzle because the branching ratio
of ${\bar B}^0\to \bar{K}^0 K^0$ will also be enhanced by them,
which will not agree with the data.

The new physics contribution, the terms proportional to $C_{11}$
and $C_{13}$ respectively, which contributes to one mode but not
to the another (precisely speaking, the contribution to the
another mode is $m_d/m_s$ suppressed) comes from the hadronic
matrix elements of Higgs penguin operators at the leading order in
$\alpha_s$. It is the contribution that gives a possibility to
make the data be account for without a conflict with all relevant
experimental measurements. The key point is if one can have a
sizable $C_{13}^{(\prime)}(\mu)$ and/or $C_{11}^{(\prime)}(\mu)$
in the 2HDM and MSSM.

Let us analyze how large $C_{13}^{(\prime)}$ and/or
$C_{11}^{(\prime)}$ are needed to fit the data. Let
\begin{eqnarray}
z&=& \sum_{p=u,c}{\lambda_p\over \lambda_t} \left( a_4^p+ra_6^p
\right),\;\;\; z_1= {\lambda_u\over \lambda_t}\;a_2, \;\;\; z_2 =
{m_s\over m_b} [h_1(C_{11}-C_{11}^\prime)+h_2
(C_{13}(\mu)-C_{13}^\prime(\mu))], \nnb %{1\over 2} {m_s\over m_b} r C_{13}(\mu),
\end{eqnarray}
%where $h_{1,2}$ are known constants arising from RQE running,
we have
\begin{eqnarray}
r_{11}\equiv\sqrt{2}\sqrt{32 \pi m_B {\rm Br}(\pi^0\pi^0)_{\rm
min} \over (G_F f_\pi F^{B\to\pi}\lambda_t)^2 \tau_B}&\leqslant&
|z-z_1 | \leqslant r_{12}\equiv\sqrt{2}\sqrt{32 \pi m_B {\rm
Br}(\pi^0\pi^0)_{\rm max} \over (G_F f_\pi
F^{B\to\pi}\lambda_t)^2 \tau_B}\nonumber\\
r_{21}\equiv\sqrt{32 \pi m_B {\rm Br}({\bar K}^0 K^0)_{\rm min}
\over (G_F f_K F^{B\to K}\lambda_t)^2 \tau_B}&\leqslant&
|z+z_2|\leqslant r_{22}\equiv \sqrt{32 \pi m_B {\rm Br}({\bar K}^0
K^0)_{\rm max} \over (G_F f_K F^{B\to K}\lambda_t)^2 \tau_B}
\end{eqnarray}
 From the data, $1.16\times 10^{-6}\leqslant{\rm
Br}(\pi^0\pi^0)\leqslant 1.74\times 10^{-6}$ and $0.71\leqslant {\rm
Br}({\bar K}^0 K^0) \leqslant 1.21 \times 10^{-6}$\cite{bauer}, we have
$r_{12}>r_{11}>r_{22}>r_{21}$. To satisfy the above two relations,
it is necessary to have
\begin{eqnarray}\label{z2z1}
|z_2+z_1| \geqslant r_{11}-r_{22}
\end{eqnarray}

\begin{figure}\vskip -1.5cm
\epsfysize=9cm \centerline{\epsffile{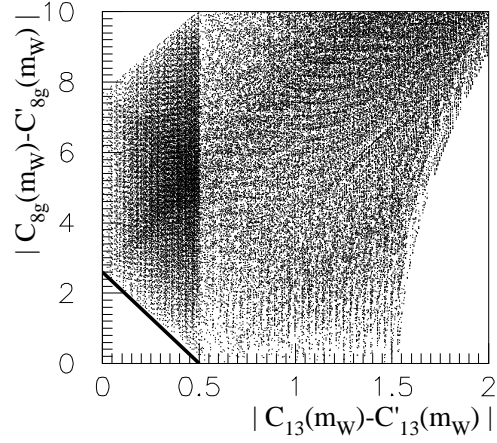}} \vspace{-1.8cm}
\caption{The correlation between
$|C_{8g}(m_W)-C_{8g}^\prime(m_W)|$ and
$|C_{13}(m_W)-C_{13}^\prime(m_W)|$. %, which can be very large(for
%example, $|C_{8g}(m_W)|\sim O(50)$ and $|C_{13}(m_W)|\sim O(10)$
%in the $1\sigma$ experiment contrainst. But such values are far
%byeond the the possible area in the new physics models, so we just
%plot the part in which we are interested.
}\label{corr}
\end{figure}

In the model I and II 2HDMs and MSSM the Wilson coefficients of
QCD penguin operators are not changed significantly, compared with
those in SM, and the Wilson coefficient of chromo-magnetic
operator can have a significant change~\cite{hkwy}. Taking the SM
values of Wilson coefficients of relevant operators but the
chromo-magnetic operator and using RQE running, we can obtain the
correlation between $|C_{8g}(m_W)-C_{8g}^\prime(m_W)|$ and
$|C_{13}(m_W)-C_{13}^\prime(m_W)|$ from $|z-z_1|\geqslant r_{11}$
and Eq.~(\ref{z2z1}), which is shown in Fig. \ref{corr} where
$C_{11}=C_{13}$ has been assumed for simplicity, without losing
the generality of discussions\footnote{In the figure
$|C_{8g}(m_W)-C_{8g}^\prime(m_W)|$ has no upper bound because we
do not impose $|z-z_1|\leqslant r_{12}$. In all models beyond SM
known so far, $|C_{8g}(m_W)-C_{8g}^\prime(m_W)|$ never reach very
large value (say, 5) when all relevant experimental constraints
are imposed. We do not need to know the upper bound for the
analysis in the paper.}.  It follows from the figure that
$|C_{8g}(m_W)-C_{8g}^\prime(m_W)|_{\rm min} = 2.6$ when
$C_{13}(m_W)-C_{13}^\prime(m_W)=0$, which reduce to
Eq.~(\ref{c8g}) in the MFV models, as it should be.

%Meanwhile, we need $|z|_{\rm min} =r_{22}+ z_2 = r_{11}-z_1=0.172$
%and $z=0.113+ 0.130 i$. Combining the Eq.(3), we have
%\begin{eqnarray}
%&&z_{\rm min} = \sum_{p=u,c} {\lambda_p\over \lambda_t} (a_4^p+ r
%a_6^p)
% = (-0.0486 - 0.00137 i) + 0.0468 C_{8g}(m_W) =0.113+ 0.130 i, \nonumber\\
%&&|C_{8g}(m_W)|_{\rm min} = 4.5.
%\end{eqnarray}

%\begin{figure}
%\subfigure[ is for the model including new operators
%$C_{11}-C_{16}$.]{
%\begin{minipage}[t]{.48\textwidth}
%\epsfysize=4cm \vspace{-1.5cm} \centering
%\includegraphics[width=7cm]{pik-final.eps} \vspace{-0.5cm}
%\end{minipage}}
%\caption{The same as Fig. 1, in models with new operators.
%}\label{independent2}
%\end{figure}

It is well-known that the experimental upper bound of branching
ratio for $B_s\to \mu^+\mu^-$ constrains severely parameters in
the MSSM and model I and II 2HDMs~\cite{bs}. Similarly, we show
that the the corresponding bound for $B_d\to \mu^+\mu^-$ implies
that the Wilson coefficients of new operators in the MSSM and
model I and II 2HDMs can not be large. The branching ratio $B_d
\rightarrow \mu^+ \mu^-$ in the 2HDM
and MSSM is given as%~\cite{bd}
\begin{eqnarray}\label{bsmu}
{\rm Br}(B_d \rightarrow \mu^+ \mu^-) &=& \frac{G_F^2
\alpha^2_{\rm em}}{64 \pi^3} m^3_{B_d} \tau_{B_d} f^2_{B_d}
|\lambda_t|^2 \sqrt{1 - 4 \widehat{m}^2}
[(1 - 4\widehat{m}^2) |C_{Q_1}(m_b) - C^\prime_{Q_1}(m_b)|^2 + \nonumber\\
&& |C_{Q_2}(m_b) - C^\prime_{Q_2}(m_b) + 2\widehat{m}(C_{10}(m_b)
- C^\prime_{10}(m_b) )|^2]
\end{eqnarray}
where $\widehat{m} = m_\mu/m_{B_d}$. In the moderate  and large
$\tan\beta$ cases the term proportional to
$(C_{10}-C_{10}^\prime)$ in Eq. (\ref{bsmu}) can be neglected. The
new CDF and D0 combined experimental upper bound of ${\rm
Br}(B_d\to \mu^+\mu^-)$ is $3.2 \times 10^{-8}$~\cite{bdsmu} at
$90\%$ confidence level.
%To translate into
%the constraint on $C_{11,13}$, we have
%\begin{eqnarray}
%\sqrt{|C_{11}(m_W)-C_{11}^\prime(m_W)|^2 +
%|C_{13}(m_W)-C_{13}^\prime(m_W)|^2}\lsim 0.047
%\end{eqnarray}
We have the constraint
\begin{eqnarray}
\sqrt{|C_{Q_1}(m_W)-C_{Q_1}^\prime(m_W)|^2 +
|C_{Q_2}(m_W)-C_{Q_2}^\prime(m_W)|^2}\lsim 2.2
\end{eqnarray}
where $C_{Q_{1,2}}^{(\prime)}$ are the Wilson coefficients of the
operators $Q_{1,2}^{(\prime)}$ which are Higgs penguin induced in
leptonic and semileptonic B decays and their definition can be
found in Refs.~\cite{dhh,hw}. By substituting the quark-Higgs
vertex for the lepton-Higgs vertex, it is straightforward to
obtain Wilson coefficients relevant to hadronic B decays in the
MSSM and model I and II 2HDMs. To translate $C_{Q_{1,2}}$ into
$C_{Q_{11,13}}$, we have $C^{(\prime)}_{Q_{11,13}}(m_W) \sim
0.037$. Then it follows from Fig.\ref{corr} that
$|C_{8g}(m_W)-C_{8g}^\prime(m_W)|$ must be larger than 2.4 in
order to resolve the puzzle.

The Wilson coefficients $C_{8g}^{(\prime)}$ in the $b\to d$
transition are constrained by Br($B\to X_d\,g$). Because there is
no data for Br of the $B\to X_d\,g$ decay and the difference
between the $B\to X_d\,g$ and $B\to X_s\,g$ decays in the SM comes
from CKM matrix elements, we assume the constraint on
$C_{8g}^{(\prime)}$ same as that from $b\to s\,g$. In the presence
of new physics a model-independent analysis gives that
$|C_{8g}(m_W)-C_{8g}^\prime(m_W)|< 2.01$ when $Br(b\to s\,g)<
9\%$~\cite{sg}. That is, $|C_{8g}(m_W)-C_{8g}^\prime(m_W)|$ can
not satisfy the condition, larger than 2.4, because of the $b\to
s\,g$ constraint. Therefore, we come to the conclusion that the
puzzle of $\bar{B}^0\to\pi^0\pi^0$ can not get resolved within
$1\sigma$ experimental bounds in the MSSM and model I and II
2HDMs.

If one does not impose the $b\to s\,g$ constraint it is possible
to resolve the puzzle in the MSSM because the Wilson coefficient
$C_{8g}^{(*)}$ can reach values larger than $2.6$ in some regions
of parameter space. We have carried out detailed numerical
calculations in the MSSM, imposing the constraints from the ${\bar
B}^0_d - \bar{B}^0_d$ system, the mass difference $\Delta M_d=
(0.509 \pm 0.004) {\rm ps}^{-1}$, mixing induced CP violation
phase angle $\beta$ measured in charmonium $B$ decays, $\sin
2\beta =0.687 \pm 0.032$~\cite{hfag}, and ${\bar B}^0\to X_d
\gamma$, in addition to the constraint from $B_d\to \mu^+\mu^-$,
however, without imposing the $b\to s\,g$ constraint.
$\delta^{LL,RR}_{13}$ and $\delta^{LR,RL}_{13}$ are constrained to
be order of $10^{-1}$ and $10^{-2}$ respectively with moderate
sparticle masses (say, $500$GeV)~\cite{masiero01,ko2002}.
%Recently there are experimental measurements on
%exclusive $B \to \rho (\omega) \gamma$. Due to large hardronic
%uncertainties theoretically, we do not consider this constraint.
%Instead, we impose
In particular, Br$(\bar{B}^0 \to X_d \gamma) \le 1 \times 10^{-5}$
extracted from exclusive $B \to \rho (\omega) \gamma$, as
advocated in Ref.~\cite{ko2002}, gives a more stringent
constraint. Br(${\bar B}^0\to X_d \gamma$) directly constrains
$|C_{7\gamma}(m_b)|^2 + |C^\prime_{7\gamma}(m_b)|^2$ at the
leading order. Due to the strong enhancement factor
$m_{\tilde{g}}/m_b$ associated with single $\delta^{LR(RL)}_{13}$
insertion term in $C^{(\prime)}_{7\gamma}(m_b)$,
$\delta^{LR(RL)}_{13}$ ($\sim 10^{-2}$) are more severely
constrained than $\delta^{LL(RR)}_{13}$. However, if the
left-right mixing of scalar bottom quark $\delta^{LR}_{33}$ is
large ($\sim 0.5$), $\delta^{LL(RR)}_{13}$ is constrained to be
order of $10^{-2}$ since the double insertion term
$\delta^{LL(RR)}_{13} \delta^{LR(LR*)}_{33}$ is also enhanced by
$m_{\tilde{g}}/m_b$. In the case of LL (RR) insertion the
precisely measured mass difference $\Delta M_d$ imposes a more
severe constraint on $\delta^{LL(RR)}_{13}$.

\begin{figure}\vskip -1cm
\epsfysize=6cm \centerline{\epsffile{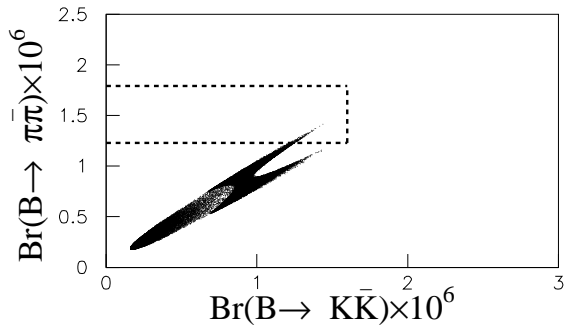}} \vspace{-1cm}
\caption{The correlation between ${\bar B}^0\to \pi^0\pi^0$ and
${\bar B}^0\to \bar KK$ in MSSM with the LL
insertion.}\label{mssm}
\end{figure}

In numerical analysis we fix $\tan\beta=10$, vary $m_{\tilde g}$
and $m_{\tilde q}$ in the region between $300$ GeV and $2$ TeV,
and the NHB masses in the ranges of $91 {\rm GeV} \leqslant m_h
\leqslant 135 {\rm GeV}, 91 {\rm GeV} \leqslant m_H \leqslant 200
{\rm GeV}$ with $m_h < m_H$ and $200 {\rm GeV} \leqslant m_A
\leqslant 240 {\rm GeV}$ for the fixed mixing angle $ \pi/2$ of
the CP even NHBs, and scan $\delta^{dAB}_{13}$ in the range
$|\delta^{dAB}_{13}| \leqslant 0.1$ for A=B and 0.05 for $A\neq B$
(A = L, R). The numerical result for the correlation between
branching ratios of ${\bar B}^0 \to \pi^0\pi^0$ and ${\bar B}^0\to
\bar K^0 K^0$ is shown in Fig.\ref{mssm} for the case of LL
insertion. Due to the combined constraints mentioned above, in
most of parameter space $C_{8g}(m_W)$ is not large enough to
accommodate the data in $1\sigma$ region. However, there are small
regions of parameter space with $x \gg 1$ (say $x \sim 40$, $x$ is
the square of the ratio between scalar quark and gluino masses)
where $C_{8g}(m_W)$ is large enough to resolve the puzzle. In the
case of both LL and RR insertion, the result is similar. In the
cases of LR, both LR and RL insertions, we also have similar
results. For $x > 1$ we have the regions with large enough
$C_{8g}(m_W)$ and the regions
%in which $C_{8g}(m_W)$ is large enough
are larger than those in the cases of LL, both LL and RR
insertions. In short, numerical results of Br show that the MSSM
can explain the puzzle within $1\sigma$ experimental bounds under
all relevant experimental constraints except that from Br($b\to
s\, g$).
%but the $b\to s\, g$ constraint.

Finally we consider the model III 2HDM~\cite{m3}. In the model III
2HDM there are tree-level flavor changing neutral currents (FCNC).
After diagonalizing the mass matrix of quark fields, the flavor
changing (FC) part of the Yukawa Lagrangian is~\cite{m3}
\begin{eqnarray}
 {\cal L}_{Y,FC}^{(III)} = \xi^{U}_{ij} \bar Q_{i,L}\tilde H_2
U_{j,R}
 + \xi^D_{ij}\bar Q_{i,L} H_2 D_{j,R} \,+\, h.c. \label{lyukfc}
\end{eqnarray}
In order to obtain naturally small FCNC one assumes the Cheng-Sher
parameterization~\cite{cs}
\begin{eqnarray}
\xi^{D}_{ij}=\lambda_{ij}\,\frac{\sqrt{m_i m_j}}{v} \label{sher}
\end{eqnarray}

Phenomenological constraints on parameters of the models have been
extensively discussed~\cite{con}. For $b\to ds\bar{s}$, the
couplings $\lambda_{bd,db,ss}$ are involved. $\lambda_{bd}$ can
reach 0.4 without a conflict with the measured mass difference
$\Delta M_{B_d}$ if the mass of pseudo-scalar Higgs boson $M_A$ is
large (say, $\sim 1 TeV$)~\cite{atwood} and it is also allowed by
the recent data for ${\bar B}^0\to \rho (\omega) \gamma$. The
constraint on $\lambda_{ss}$ from the analysis on ${\bar
B}^0_s-\bar{B}^0_s$ shows that $\lambda_{ss}$ can reach
$O(100)$~\cite{hl} which means that the coupling of Higgs to $s$
quark is $O(10^{-2})$. It should be emphasized that the constraint
from $B_d \to \mu^+\mu^-$ is irrelevant in the model III 2HDM
because the decay involves $\lambda_{\mu\mu}$ besides
$\lambda_{bd}$ and $\lambda_{\mu\mu}$ has no relation to
$\lambda_{ss}$, which is different from the MSSM and 2HDMs I and
II.

In numerical calculations, we use $m_h=120{\rm GeV}, %m_A=1000{\rm GeV},
m_d=6{\rm MeV}, \lambda_{bd}=0.3, \lambda_{ss}=150$ and
consequently obtain $C_{13}(m_W)= 0.41$. Corresponding to this
value, $|C_{8g}(m_W)|\geqslant 0.6$, which can be satisfied under
all relevant constraints. The numerical result for the values of
parameters given above shows that
\begin{eqnarray} Br({\bar B}^0\to \pi^0\pi^0)= 1.3\times
10^{-6},~~~~~~~~ Br({\bar B}^0\to {\bar K}^0K^0)=0.9\times
10^{-6}. \nonumber \end{eqnarray} That is, the data of branching
ratios of ${\bar B}^0\to \pi^0\pi^0$ and ${\bar B}^0\to {\bar
K}^0K^0$ are accounted for, as expected. At the same time, we have
checked that the NP contribution to Br(${\bar B}^0\to \pi^+\pi^-$)
is very small and negligible.

In conclusion, we have analyzed the ${\bar B}^0\to \pi^0\pi^0$
puzzle in three kinds of models beyond the SM. In the analysis
$1\sigma$ experimental bounds and $60\%$ theoretical uncertainty
which mainly comes from the input of non-perturbative parameters
have been taken into account. It is shown that the minimal flavor
violation models, the minimal supersymmetric standard model, and
the two Higgs doublet models I and II can not give an explanation
of the ${\bar B}^0\to \pi^0\pi^0$ puzzle within $1 \sigma$
experimental bounds when all relevant experimental constraints are
imposed and the model III 2HDM can explain the puzzle without a
conflict with other experimental measurements. Therefore, if the
data of Br for ${\bar B}^0\to \pi^0\pi^0$ and ${\bar B}^0\to {\bar
K}^0K^0$ persist in the future, the MFV models, MSSM and model I,
II 2HDMs will be excluded within $1 \sigma$ experimental bounds
and the model III 2HDM will be survived to resolve the puzzle. As
it is obvious, the analysis depends on the estimate of theoretical
uncertainty. Our estimate comes from the uncertainties of input
parameters (form factors, distribution amplitudes, CKM matrix
elements, etc.) as well as the error estimate of neglecting
electro-weak penguin and annihilation contributions. If the
uncertainty were $70\%$, $|C_{8g}(m_W)-C_{8g}^\prime(m_W)|_{min}$,
which should be reached in a model in order to account for the
data, would be $2.2$ in the MFV models and $2.0$ in the MSSM
respectively so that the MFV models could not give an explanation
of the data and the MSSM could. We have also analyzed the puzzle
in the case without imposing the $b\to s\,g$ constraint in the
MSSM. For all kinds of insertions there are regions of parameter
space where the puzzle can be resolved within $1 \sigma$
experimental bounds. We expect that similar effects appear in
decays with $PV$ final states, $B\to \pi^0 \rho^0$ and $B\to
\bar{K}^0 K^{0*}$.
\section*{Acknowledgement}
The work was supported in part by the Natural Science Foundation
of China (NSFC), grant 10435040, grant 90503002,
and grant 10225522.

\end{document}